\input{aipcheck}

\documentclass[
    ,final            % use final for the camera ready runs
  ]
  {aipproc}

\layoutstyle{8x11single}

\usepackage{amssymb, amsmath}
\usepackage{graphicx}% Include figure files
\usepackage{bm}% bold math
\usepackage{braket}
\usepackage{natbib}

\usepackage{textcomp}
\begin{document}

\title{Superradiance in a Two-Channel Quantum Wire}



\author{A.~Tayebi}{
  address={Department of Physics and Astronomy, Michigan State University, East Lansing, Michigan 48824, USA}
  ,altaddress={Department of Electrical and Computer Engineering, College of Engineering, Michigan State University, East Lansing, Michigan 48824, USA} % additional visiting address
}

\author{V.~Zelevinsky}{
  address={Department of Physics and Astronomy, Michigan State University, East Lansing, Michigan 48824, USA}
  ,altaddress={National Superconducting Cyclotron Laboratory, Michigan State University, East Lansing, Michigan 48824, USA} % additional visiting address
}

\begin{abstract}
 A one-dimensional, two-channel quantum wire is studied in the effective non-Hermitian Hamiltonian framework. Analytical expressions are derived for the band structure of the isolated wire. Quantum states and transport properties of the wire coupled to two ideal leads at the edges are studied in detail. The width distribution of the quasistationary states varies as a function of the coupling strength to the environment. At weak coupling, all the eigenenergies uniformly acquire small widths. The picture changes entirely at strong coupling, a certain number of states (``super-radiant'') are greatly broadened, while the rest remain long-lived states, a  pure quantum mechanical effect as a consequence of quantum interference. The transition between the two regimes
%, where the system undergoes a sharp width re-distribution, the so called super-radiance %transition,
greatly influences the transport properties of the system. The maximum transmission through the wire occurs at the super-radiance transition. We consider also a realistic situation with
energy-dependent coupling to the continuum due to the existence of decay threshold where  super-radiance still  plays a significant role in transport properties of the system.
\end{abstract}

\maketitle

%%%%%%%%%%%%%%%%%%%%%%%%%%%%%%%%%%%%%%%%%%%%
%% MAINMATTER
%%%%%%%%%%%%%%%%%%%%%%%%%%%%%%%%%%%%%%%%%%%%

\section{Introduction}

The studies of open quantum systems range from loosely bound nuclei to quantum information and teleportation devices. Quantum transport through open quantum systems has recently gained significant attention. The advent of nano-fabrication technology has opened new doors for studying quantum transport, in particular electronic transport. This area of research has further expanded with the inclusion of spin and thermodynamic degrees of freedom in the newborn fields of spitronics and spin-caloritronics. In nuclear physics, especially for short-lived
nuclei currently in the center of interest, the structure and reactions are greatly influenced by the proximity of the continuum. In all such applications we have to deal with an open quantum system where the interaction between the bound states and the environment is not only taken into account but frequently presents the main object of study.

An appropriate mathematical framework for studying open quantum systems is provided by the effective non-Hermitian Hamiltonian, which was first introduced by Feshbach \cite{Feshbach} in the context of nuclear physics \cite{MW,SZNPA}. The applicability of the framework goes beyond nuclear physics \cite{AZ11}, it ranges from quantum biology to electronic transport in mesoscopic physics. Recently \cite{Cel-bact}, a quantum biological switch was demonstrated using the same mathematical formalism. Transport properties in a periodic and disordered chains with symmetric \cite{celardo09} and asymmetric \cite{Celardo10} coupling to the leads and signal transmission properties of a two-level atom, a qubit, inserted into a periodic chain \cite{Tayebi} are typical examples of numerous applications where the effective non-Hermitian Hamiltonian approach was shown to be a powerful tool.

 A common feature for all open systems is the appearance of short-lived states at sufficiently strong coupling to the external world. These states are called super-radiant, analogous to the Dicke super-radiant states in quantum optics \cite{dicke54}. Super-radiant states are broad resonances that occur as a result of the collectivization of the intrinsic states of the system due to their coupling through the common decay channels. This collective phenomenon was experimentally demonstrated in \cite{Scheibner} by using an ensemble of quantum dots. At the same time, the remaining (``trapped'') states become very long-lived. Varying the effective coupling one can envision interesting possibilities to regulate quantum networks.

In what follows, we study super-radiance and its effect on transport properties in a quantum wire prototype: a one-dimensional chain of two level systems where the edges are coupled to the outside world presented by a continuum of channels. The two-channel wire has served as a model in \cite{Dagotto} where superconductivity in ladder structures was studied and in \cite{nakamura07} where a two-dimensional chain was coupled to an adatom. This specific system was not earlier analyzed from the viewpoint of super-radiance.

The paper is organized as follows. In section \MakeUppercase{\romannumeral 2} we consider the isolated wire, a closed chain of $N$ identical two-level cells with the nearest neighbor hopping interaction between the ground and excited states of neighboring cells. A closed form solution describes the two energy bands of the system. Several symmetries associated with the eigenvectors of the system are explained as well. A brief description of the effective non-Hermitian Hamiltonian is provided in section \MakeUppercase{\romannumeral 3}. The system is then opened through the coupling of the left and right edges of the chain to the continuum. As a result of the coupling the eigenenergies acquire an imaginary part. The evolution of the complex energies as a function of the coupling strength is studied in detail. It is shown that the width is uniformly distributed at weak coupling. However, the picture changes at strong couplings: the states are not uniformly broadened. The super-radiant states, their number being equal to the number of intrinsic doorway states directly coupled to the continuum, are greatly broadened while the remaining states acquire only small widths. The transport properties of the system are studied in section \MakeUppercase{\romannumeral 4}. It is shown that the transition between the weak and the strong coupling regimes, the so-called super-radiance transition, greatly influences the scattering properties of the system, particularly the transmission coefficient. Energy-dependent coupling to the continuum due to the presence of the decay threshold is considered in section \MakeUppercase{\romannumeral 5}. It is shown that super-radiance plays a significant role, even when the energy levels are in proximity to their cutoff thresholds. The paper concludes in section \MakeUppercase{\romannumeral 6}.

\section{The Isolated Wire}

\begin{figure}[!ht]
%\begin{center} height=.3\textheight
\includegraphics[keepaspectratio = true, width = 5.0 in, clip = true]{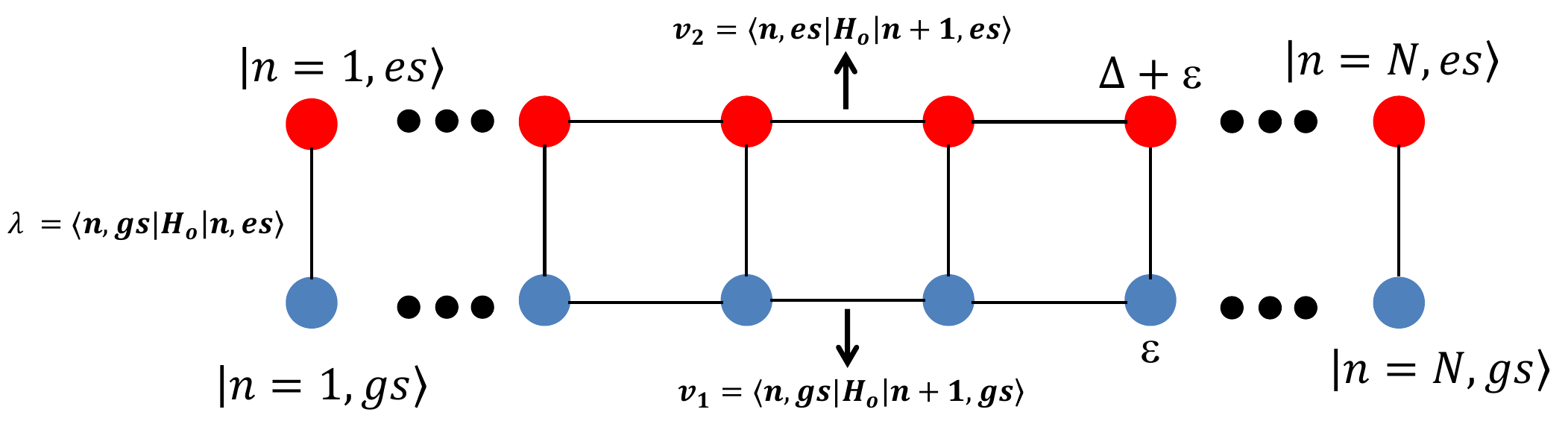}
%\end{center}
\caption{(Color online). \small{Schematic of the two-channel wire.}} \label{fig:001}
\end{figure}

The wire of the model consists of a periodic chain of $N$ identical two-level cells. The energy levels of the ground state $|n,g\rangle$ and the excited state $|n,e\rangle$ of each cell are $\epsilon$ and $\epsilon+\Delta$, respectively. The matrix element for excitation between the ground and excited state of each cell is denoted $\lambda$, and the hopping amplitudse between the neighboring ground and excited states are $v_{1}$ and $v_{2}$, respectively. Fig. \ref{fig:001} shows a schematic of the system under study.

The Hamiltonian of the closed chain is

\begin{align}\label{eqn1}
H_{0} =  \sum_{n=1}^{N} & \Big\{ \frac{\epsilon}{2}\ket{n,g}\bra{n,g}+ \frac{\epsilon+\Delta}{2}\ket{n,e}\bra{n,e}\nonumber \\
  & +\lambda\ket{n,g}\bra{n,e}+ v_{1}\ket{n,g}\bra{n-1,g}\nonumber  \\
& +v_{2}\ket{n,e}\bra{n-1,e}+h.c. \Big\} .
\end{align}

In order to diagonalize the Hamiltonian a stationary state with energy $E$ is considered
as a superposition of localized states,

\begin{equation}\label{eqn2}
\ket{q}=\sum_{n=1}^{N}\left[ a_{n}(E)\ket{n,g} + c_{n}(E)\ket{n,e}\right].
\end{equation}

%Plugging the stationary state into
The Schr\"{o}dinger equation with the Hamiltonian given in eq. (\ref{eqn1}) determines a set of two coupled linear recurrence equations for the amplitudes $a_{n}$ and $c_{n}$:

\begin{equation}\label{eqn3}
(\epsilon+\Delta)c_{n}+\lambda a_{n}+v_{2}(c_{n-1}+c_{n+1})=Ec_{n}, \quad n=1,2,..,N,
\end{equation}

\begin{equation}\label{eqn4}
\epsilon a_{n}+\lambda c_{n}+v_{1}(a_{n-1}+a_{n+1})=Ea_{n}, \quad n=1,2,..,N.
\end{equation}

Ordinary decoupling of the equations would lead to a fifth order recurrence relation. However, the decoupled equations can be simplified with the realization of a symmetry resulting from the parallel ladder structure of the two-channel wire:
\begin{equation}\label{eqn5}
c_{n}=m(E) a_{n}.
\end{equation}
Rewriting eq. (\ref{eqn3}) using eq. (\ref{eqn5}) results in a second order difference equation:
\begin{equation}\label{eqn6}
\left\{ (\epsilon + \Delta)+\frac{\lambda}{m}\right\}c_{n}+v_{2}(c_{n-1}+c_{n+1})=E c_{n}.
\end{equation}
The solution to this equation, after applying the boundary condition $c_{n=0}=c_{n=N+1}=0$, is a standing wave with the quasimomentum $q$,
\begin{equation}\label{eqn7}
c_{n}=A(-1)^n \sin(n\phi_{q}),   \quad   \phi_{q}=\frac{\pi q}{N+1}.
\end{equation}

One can find the energy levels of the system using eqs. (\ref{eqn5}), (\ref{eqn7}), and the original set of the two recurrence relations. The eigenenergies are solutions to the following quadratic equation:
\begin{equation}\label{eqn8}
(\epsilon+\Delta)+\frac{\lambda^2}{E-\epsilon+2v_{1}\cos\phi_{q}}-2v_{2}\cos\phi_{q}=E.
\end{equation}
Hence the energy levels are
\begin{eqnarray}\label{eqn9}
  & E_{q}^{\pm} =\, \frac{1}{2}\,\left[-\alpha_{q} \pm \sqrt{\alpha_{q}^{2}-4\beta_{q}}\right],   \nonumber \\ &\alpha_{q}= 2v_{1}\cos\phi_{q}-\epsilon+2v_{2}\cos\phi_{q}-(\epsilon+\Delta), \nonumber \\
&\beta_{q}= -2\epsilon v_{2}\cos\phi_{q}+4v_{1}v_{2}\cos^{2}\phi_{q} \nonumber\\
&-\lambda^{2}+(\epsilon+\Delta)(\epsilon-2v_{1}\cos\phi_{q}).
\end{eqnarray}
The coefficient $m(E)$ is found  simultaneously:
\begin{equation}\label{eqn10}
m_{q}^{\pm}(E)=\frac{E_{q}^{\pm}-\epsilon+2v_{1}\cos\phi_{q}}{\lambda}.
\end{equation}
Finally, the normalization of the eigenvectors results in
\begin{eqnarray}\label{eqn11}
& c_{n}^{q,\pm}(E)=(-1)^{n}\sqrt{\frac{2}{N+1}}\frac{m_{q}^{\pm}}{\sqrt{1+(m_{q}^{\pm})^2}}\sin(n\phi_{q}),\nonumber\\
& a_{n}^{q,\pm}(E)=\frac{c_{n}^{q,\pm}(E)}{m_{q}^{\pm}},   \quad q=1,2,...,N.
\end{eqnarray}

\begin{figure}[!ht]
%\begin{center}
\includegraphics[keepaspectratio = true, width = 3.6 in, clip = true]{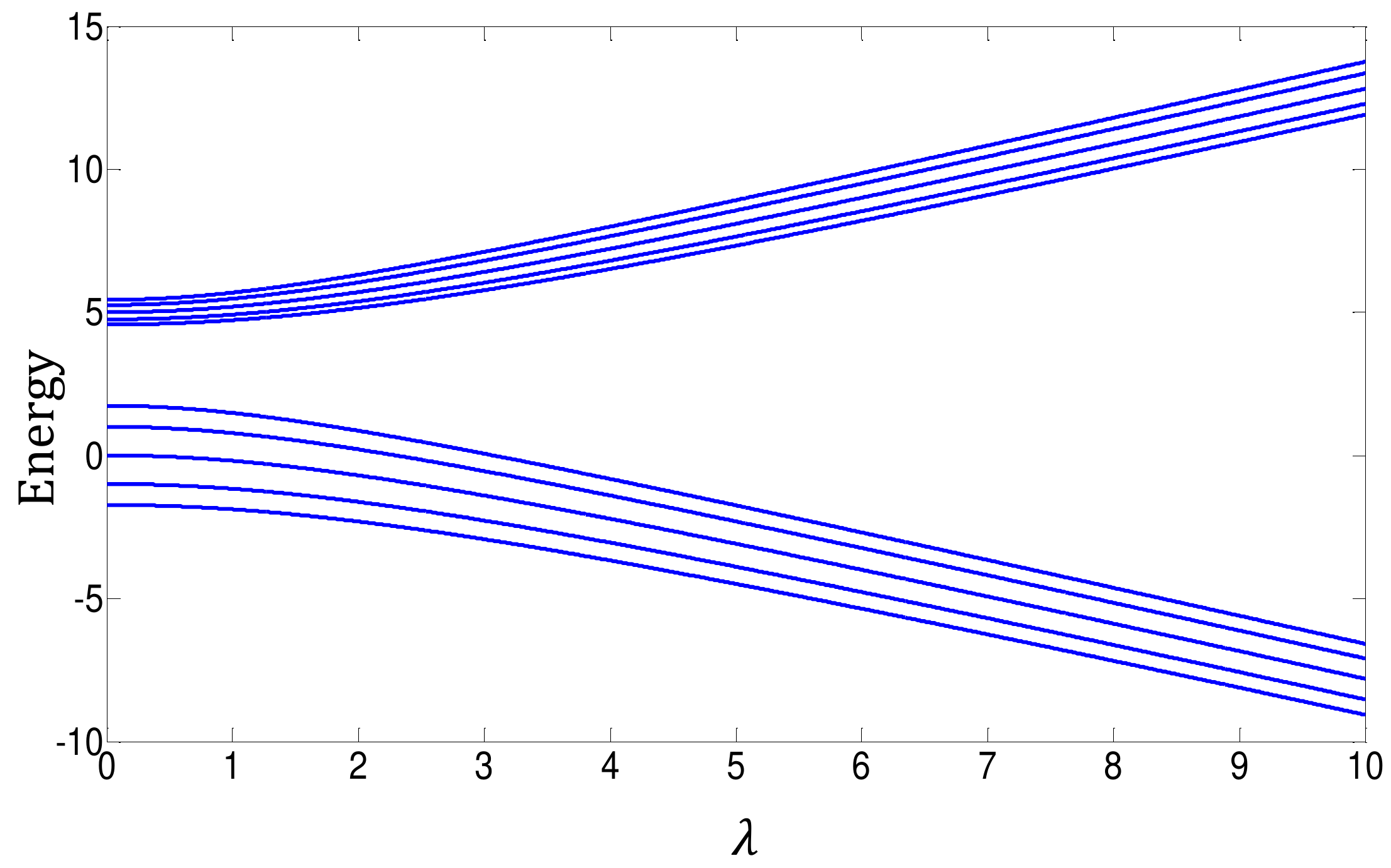}
%\end{center}
\caption{(Color online). \small{Energy levels of a two-band wire as a function of the coupling between the ground and excited states of each cell, the parameters of the wire are: $N=5, \epsilon=0, \Delta=5, v_{1}=1$ and $v_{2}=2$.}} \label{fig:002}
\end{figure}

Similar to \cite{Tayebi}, the eigenvectors can be classified into two categories: symmetric and anti-symmetric states with respect to the center of the wire. From the equation for the energy levels, eq. (\ref{eqn8}), it is easy to establish
an extra symmetry associated with the eigenvectors,
%belongs to this system. , using the general properties of quadratic equations, it can be shown that:
\begin{equation}\label{eqn12}
E_{q}^{+}E_{q}^{-}-(\epsilon+2v_{1}\cos\phi_{q})(E_{q}^{+}+E_{q}^{-})+(\epsilon+2v_{1}\cos\phi_{q})^{2}=-\lambda^{2},
\end{equation}
and use eq. (\ref{eqn12}) to show that
\begin{equation}\label{eqn13}
\frac{|m_{q}^{-}|}{\sqrt{1+(m_{q}^{-})^{2}}}=\frac{1}{\sqrt{1+(m_{q}^{+})^{2}}}.
\end{equation}
As a result, for a fixed state $q$,
\begin{equation}\label{eqn14}
|c_{n}^{q,\pm}|=|a_{n}^{q,\mp}|, \quad n=1,2,...,N.
\end{equation}
Fig. \ref{fig:002} shows the energy levels of the two bands as functions of the coupling between the ground and the excited states ($\lambda$) for a system with $N=5$. As expected, the energy plot is similar to a typical semiconductor band structure; the bands repel each other when the excitation amplitude $\lambda$ grows.

\section{Coupling to the Continuum and Super-radiance}

Now we open the system coupling the excited states at the edges to the ideal leads which are characterized by a continuum of states. With the help of projector operators, the continuum space could be eliminated. As a result, one is left with an effective non-Hermitian Hamiltonian that acts only in the internal space of the closed system. A detailed derivation
and discussion of the effective Hamiltonian are provided in \cite{AZ11}. The effective Hamiltonian of the open system
can be presented as
\begin{equation}\label{eqn15}
{\cal H}_{eff}=H_{0}-\,\frac{i}{2}W,
\end{equation}
where $H_{0}$ is the Hamiltonian of the isolated wire discussed in the previous section. The second item in the effective Hamiltonian is an anti-Hermitian term which introduces finite lifetimes to the intrinsic states of the open system. The matrix elements of $W$ without loss of generality can be written in terms of the coupling amplitudes $A_{i}^{c}$ between the intrinsic state $\ket{i}$ and the continuum state $\ket{c;E}$,
\begin{equation}\label{eqn16}
W_{ij}=\sum_{c;\,{\rm open}}A_{i}^{c}(E)A_{j}^{c}(E),
\end{equation}
where sum includes all channels in the continuum opened at a given energy. Therefore the coupling amplitudes are in general energy-dependent and vanish below a certain threshold. This energy dependence becomes important for running energies close to the threshold, otherwise the amplitudes are smooth functions of the energy and energy dependency can be ignored.

We assume that the left, $\ket{n=1,e}$, and the right, $\ket{n=N,e}$, edge state are each coupled to one channel in the continuum. Correspondingly, the squares of the coupling amplitudes determine the partial widths $\gamma^{L}$ and
$\gamma^{R}$ of the left and right edge states decoupled from the chain,
\begin{eqnarray}\label{eqn17}
A^{c={\rm left}}_{n=1,e}=\sqrt{\gamma^{L}}, \nonumber \\
A^{c={\rm right}}_{n=N,e}=\sqrt{\gamma^{R}}.
\end{eqnarray}
In the basis of unperturbed chain states, the operator $W$ has only two matrix elements,
\begin{equation}\label{eqn18}
W_{\ket{n=1,e},\ket{n=1,e}}=\gamma^{L}, \quad W_{\ket{n=N,e},\ket{n=N,e}}=\gamma^{R}.
\end{equation}

In order to diagonalize the effective Hamiltonian, $W$ is transformed into the eigenvector basis of the closed system,  $\ket{q,\pm}$.
\begin{equation}\label{eqn19}
W_{\ket{q,\pm},\ket{q',\pm}}=\gamma^{L} c_{n=1}^{q,\pm} c_{n=1}^{q',\pm} + \gamma^{R} c_{n=N}^{q,\pm} c_{n=N}^{q',\pm},
\end{equation}
where $c_{n}^{q,\pm}$ are given in eq. (\ref{eqn11}). The effective Hamiltonian can now be diagonalized which leads to the following secular equation for the complex energies $\varepsilon$:
\begin{equation}\label{eqn20}
\Omega(\varepsilon) \equiv 1+iP_{+}(\varepsilon) (\gamma^{L}+\gamma^{R}) +  \gamma^{L} \gamma^{R}(P_{-}^{2}(\varepsilon)-P_{+}^{2}(\varepsilon))=0,
\end{equation}
where $P_{+}(\varepsilon)$ and $P_{-}(\varepsilon)$ are defined as
\begin{eqnarray}\label{eqn21}
P_{+}(\varepsilon)=\frac{1}{N+1} \sum_{q,\,\pm} (1)^{q} \frac{\sin^{2}\phi_{q} m_{q}^{\pm}}{(\varepsilon-E_{q}^{\pm})(1+(m_{q}^{\pm})^{2})} \nonumber
\end{eqnarray}
and
\begin{eqnarray}\label{eqn22}
P_{-}(\varepsilon)=\frac{1}{N+1} \sum_{q,\,\pm} (-1)^{q} \frac{\sin^{2}\phi_{q} m_{q}^{\pm}}{(\varepsilon-E_{q}^{\pm})(1+(m_{q}^{\pm})^{2})}. \nonumber
\end{eqnarray}

\begin{figure}[!ht]
%\begin{center}
\includegraphics[keepaspectratio = true, width = 4.0 in, clip = true]{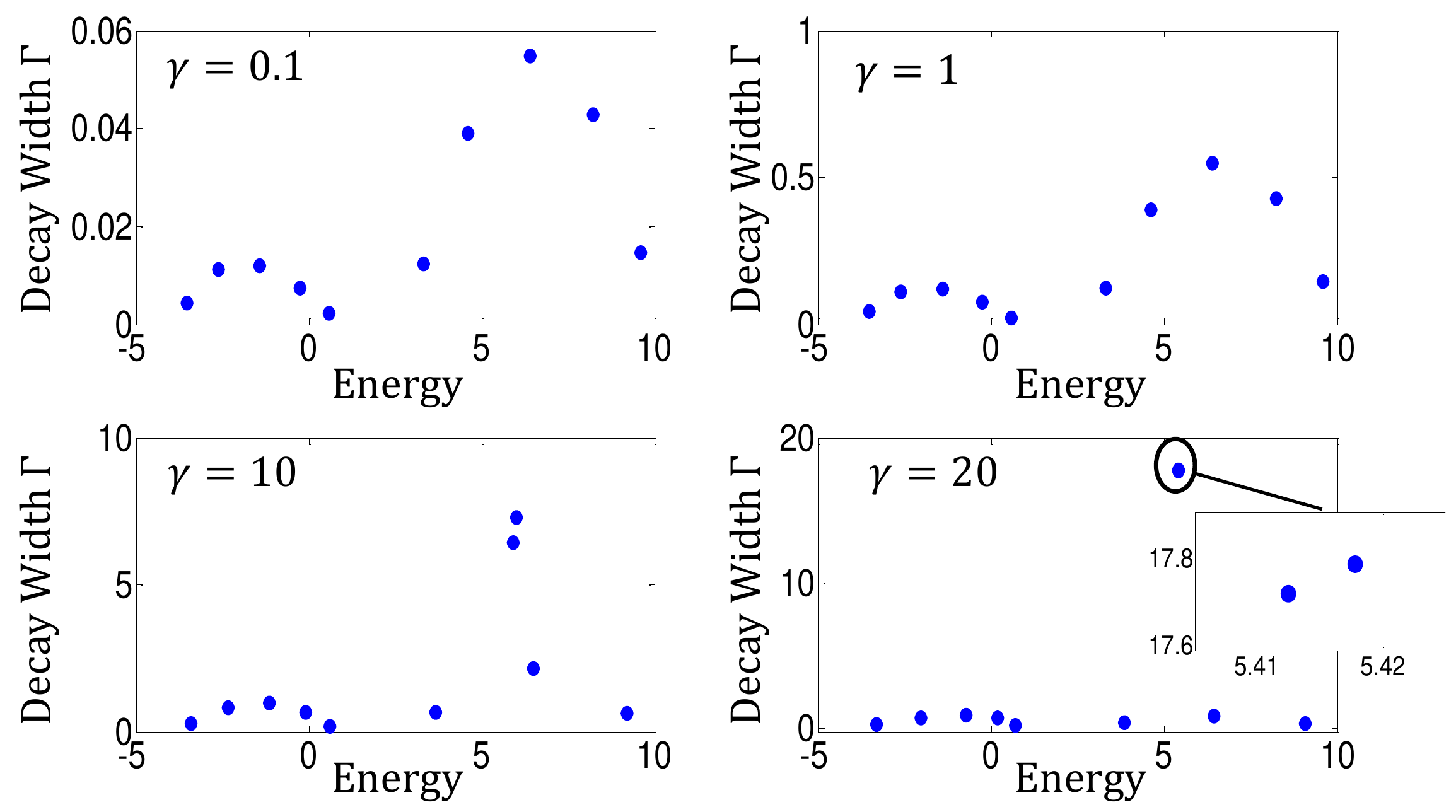}
%\end{center}
\caption{(Color online). \small{Complex eigenenergies for different values of the coupling strength to the continuum, $\gamma$. The parameters of the wire are: $N=5, \epsilon=0, \Delta=5, \lambda=3, v_{1}=1$ and $v_{2}=2$.}} \label{fig:003}
\end{figure}

Fig. \ref{fig:003} shows the resonance complex energies $E-(i/2)\Gamma$ of our system with $N$=5, when the two ends are symmetrically coupled to the continuum, $\gamma^{L}=\gamma^{R}=\gamma$. At weak coupling, when $\gamma$=0.1, all eigenenergies gain small decay width $\Gamma$ due to interaction with the environment. Further increase of $\gamma$ initially results in the increase of the decay widths for all resonances as illustrated in the upper right panel
($\gamma=1$). The complex pole dynamics changes at larger values of $\gamma$ (see \cite{AZ11} and references therein).
At a critical point, when $\gamma/D\simeq$1, where $D$ is the mean level spacing of the energy levels of $H_{0}$, the system undergoes a super-radiance transition. Beyond this point the decay widths do not monotonically increase with the increase of $\gamma$. In fact, only some states (super-radiant) continue to increase their decay widths while the remaining states become long-lived (trapped) states. This phenomenon is illustrated in the lower left panel of Fig. \ref{fig:003}, where $\gamma$=10 and the two super-radiant states, in accordance with the number of open channels, have obtained larger decay widths compared to the other states. The picture remains similar at larger $\gamma$. At $\gamma$=20, the lower right panel, the two overlapping super-radiant states are shown in the smaller window. Note that the number of super-radiant states is always equal to the number of continuum channels coupled to the wire; it is also important to
notice the different vertical scales on these panels.

\begin{figure}[!ht]
\includegraphics[keepaspectratio = true, width = 4.2 in, clip = true]{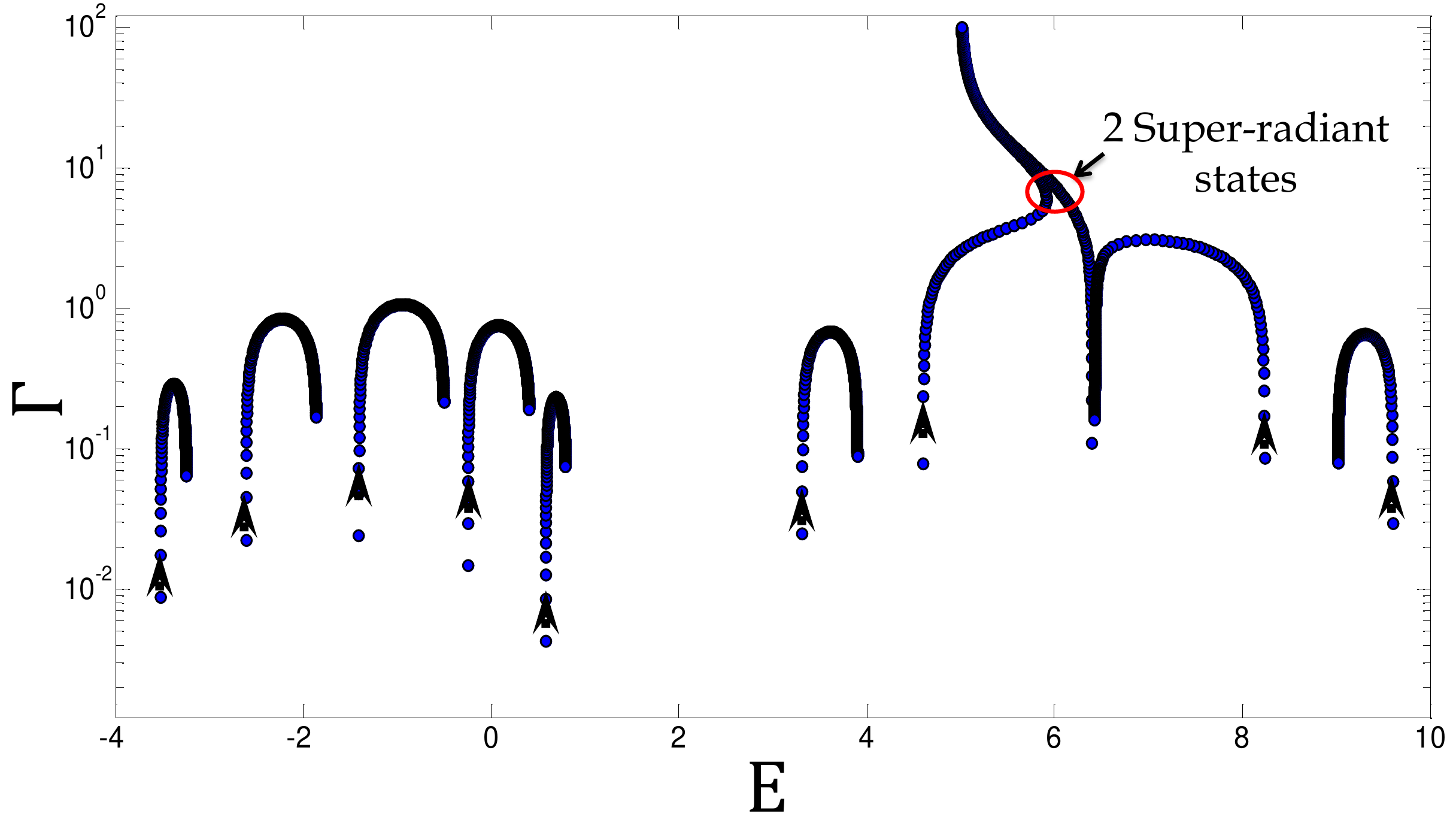}
\caption{(Color online). \small{Evolution of the energies in the complex plane. Up to the critical point all the decay widths increase as $\gamma$ increases. A super-radiant transition occurs as $\gamma$ exceeds the critical value. After passing the critical point, the decay width of the two super-radiant states increases with the increase of $\gamma$, whereas the decay width of the other eight states decreases. The parameters of the wire are similar to the one in Fig . \ref{fig:003}. The arrows show the direction of the evolution as $\gamma$ varies from 0 to 100. }} \label{fig:004}
\end{figure}

It is instructive to consider the evolution of the resonances in the complex plane as $\gamma$ increases. This is depicted in Fig. \ref{fig:004}.

\section{Transmission through the wire}

Scattering properties of the two-channel quantum wire, in particular transmission of a signal through the wire at weak, intermediate and strong coupling regimes, are considered in this section. This concerns the features which could be the most significant for any practical application.

The full Hermitian formalism provides a suitable platform for studying scattering properties of open quantum systems. In general, the transmission coefficient of a process $a \rightarrow b$, where a particle enters from the  channel $a$, which is coupled to the intrinsic state of the closed system $\ket{i}$ with the coupling amplitude $A_{i}^{a}$, propagates
through the system with possible virtual excursions to the continuum and back, and finally exits from the intrinsic state $\ket{j}$ to the decay channel $b$ with the amplitude $A^{b}_{j}$, is given by \cite{SZAnn}:
\begin{equation}\label{eqn23}
T^{a\rightarrow b}(E)=\mid Z^{a\rightarrow b}(E) \mid ^{2}.
\end{equation}
Here the amplitude $Z^{a\rightarrow b}(E)$ is given by:
\begin{equation}\label{eqn24}
Z^{a\rightarrow b}(E)=  \sum_{i,\,j} A^{b\ast}_{i} \left(\frac{1}{E-{\cal H}_{eff}}\right)_{ij}A^{a}_{j}.
\end{equation}
The sum in eq. (\ref{eqn24}) runs over all possible paths that the reaction could proceed through. In our case of the two-channel wire, we assume that the signal enters from the left and exits from the right edge of the wire or vice versa, thus the continuum channels are only coupled to the two edge cells with amplitudes given in eq. (\ref{eqn16}). Of course,
one can also look at the reflection channels; the reflectivity coefficient is $R=1-T$, since the form of the anti-Hermitian part (\ref{eqn16}) preserves the unitarity of the scattering matrix.

\begin{figure}[!ht]
\includegraphics[keepaspectratio = true, width = 4.4 in, clip = true]{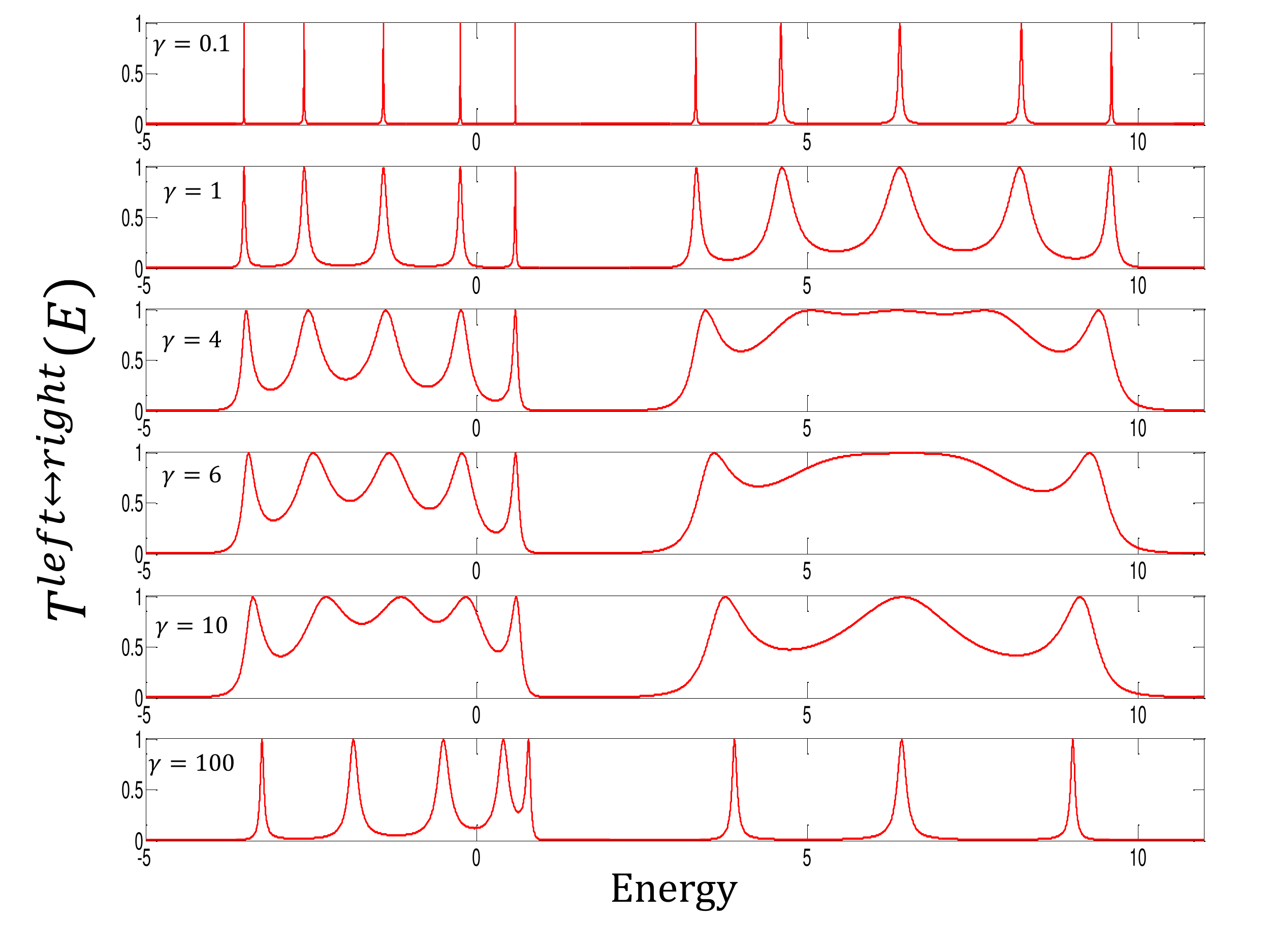}
\caption{(Color online). \small{Transmission through the two-channel wire.  The parameters of the wire are similar to those in Fig . \ref{fig:003}.}} \label{fig:005}
\end{figure}

In order to study the effect of super-radiance on the transport properties of the wire, the transmission coefficient
$T^{L\leftrightarrow R}(E)$ is depicted for different values of $\gamma$, Fig. \ref{fig:005}. At small $\gamma$, the perfect transmission, $T=1$, occurs at narrow resonances. With increasing the coupling, the resonances broaden and start to overlap. At the super-radiance transition, the maximum transmission is achieved as a result of the broadening of the states, the case of $\gamma$=4. With further increase of $\gamma$ there are only three resonances left in the upper band, the two super-radiant states are out of the picture since they are extremely broaden ($\gamma$=100). Note that the resonances become narrow as the trapped states lose their widths.

\section{Energy-dependent coupling}

In previous sections it was assumed that the matrix elements of the effective Hamiltonian do not depend on the running energy $E$. The energy dependence appeared only due to the resonance denominators. Here we briefly address a more realistic situation with energy-dependent coupling to the continuum. This dependence arises inevitably when the energy of an experiment is in the proximity of thresholds of the continuum channels (separation energy in nuclei or the work function in solids). In transmission of electrons through a crystallic chain, the energy dependence can be regulated by the applied voltage. As a result, the assumption of coupling amplitudes being constants or smooth function of energy is not valid and one should consider the energy dependency of the decay amplitudes. When a simple case of the $s$-wave form dependence is considered, the coupling amplitudes can be taken as
\begin{eqnarray}\label{eqn25}
A^{L}_{n=1,e}(E)=\alpha^{L} \Theta(E-E^{L}_{c}) \sqrt{E-E^{L}_{c}} \nonumber \\
A^{R}_{n=N,e}(E)=\alpha^{R} \Theta(E-E^{R}_{c}) \sqrt{E-E^{R}_{c}},
\end{eqnarray}
where $\alpha^{L}$ and $\alpha^{R}$ are the strengths of the coupling to the left and to the right edge, respectively. In general these coefficients could be different; similar to the previous section we assume a symmetric coupling to the continuum, hence $\alpha^{L}=\alpha^{R}=\alpha$. $E^{L}_{c}$ and $E^{R}_{c}$ are the cutoff energies of the continuum coupling also assumed to be the same, $E_{c}$, for both channels. $\Theta$ is the step function, which guarantees that the system is decoupled from the environment at energy below threshold.

\begin{figure}[!ht]
\includegraphics[keepaspectratio = true, width = 4.0 in, clip = true]{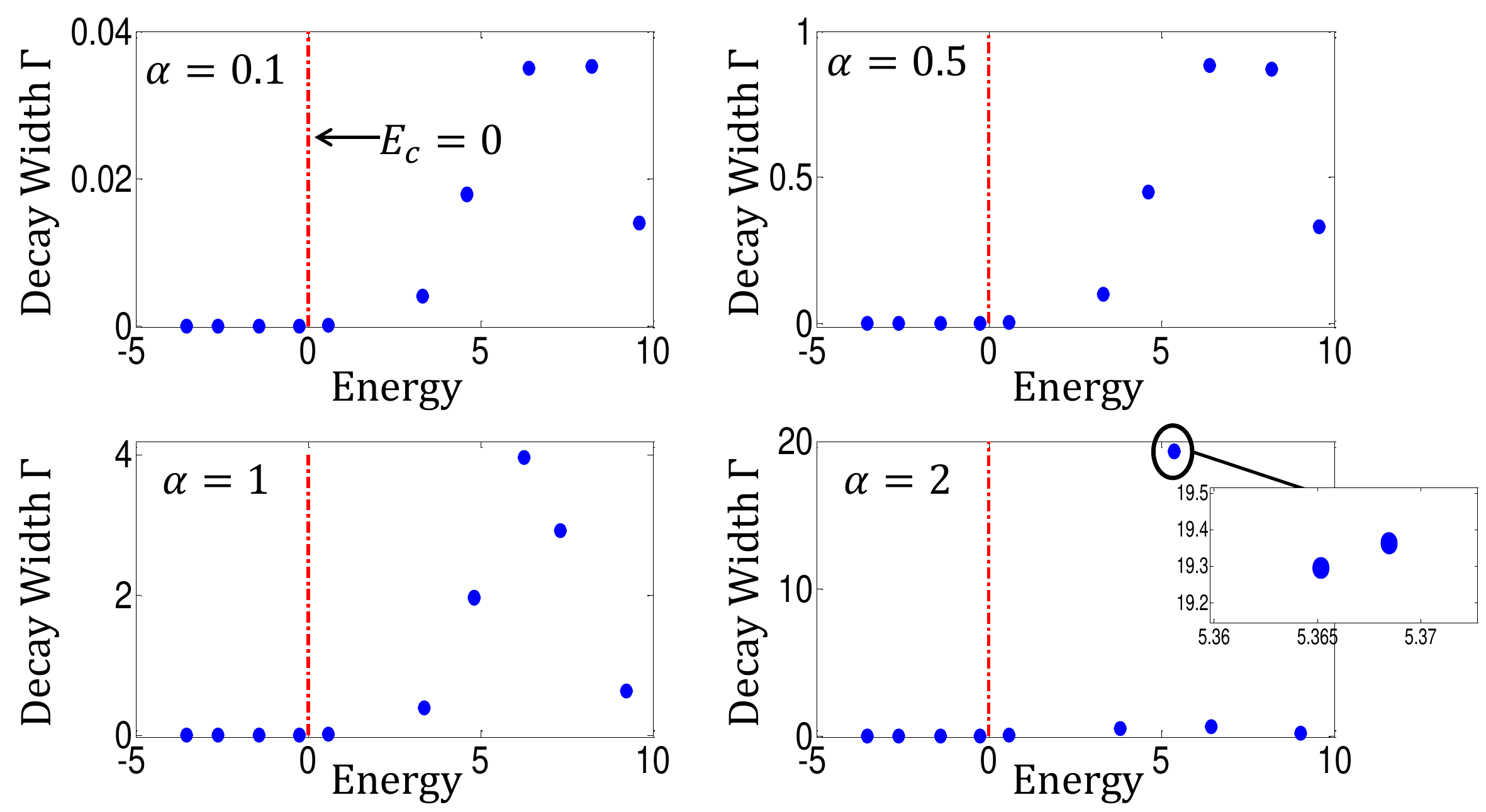}
\caption{(Color online). \small{Complex energies in the case of energy-dependent couplings for different values of the coupling strength, $\alpha$. The cutoff threshold is shown with a vertical line, $E_{c}$=0. The wire parameters are the same as in Fig . \ref{fig:003}. }} \label{fig:006}
\end{figure}

Fig. \ref{fig:006} shows the complex energies of the same system when the cutoff threshold $E_{c}$ is approximately located in between the two bands of the isolated wire. Note that the super-radiant transition still occurs as the coupling strength $\alpha$ increases. The four states with energy below threshold do not acquire any width being decoupled from the environment. The four trapped states above threshold initially gain width as $\alpha$ increases, however, beyond the super-radiant transition, when $\alpha$=2, they become long-lived states. On the lower right panel the two overlapping super-radiant states with the large decay width are shown in a smaller window.

\begin{figure}[!ht]
\includegraphics[keepaspectratio = true, width = 4.4 in, clip = true]{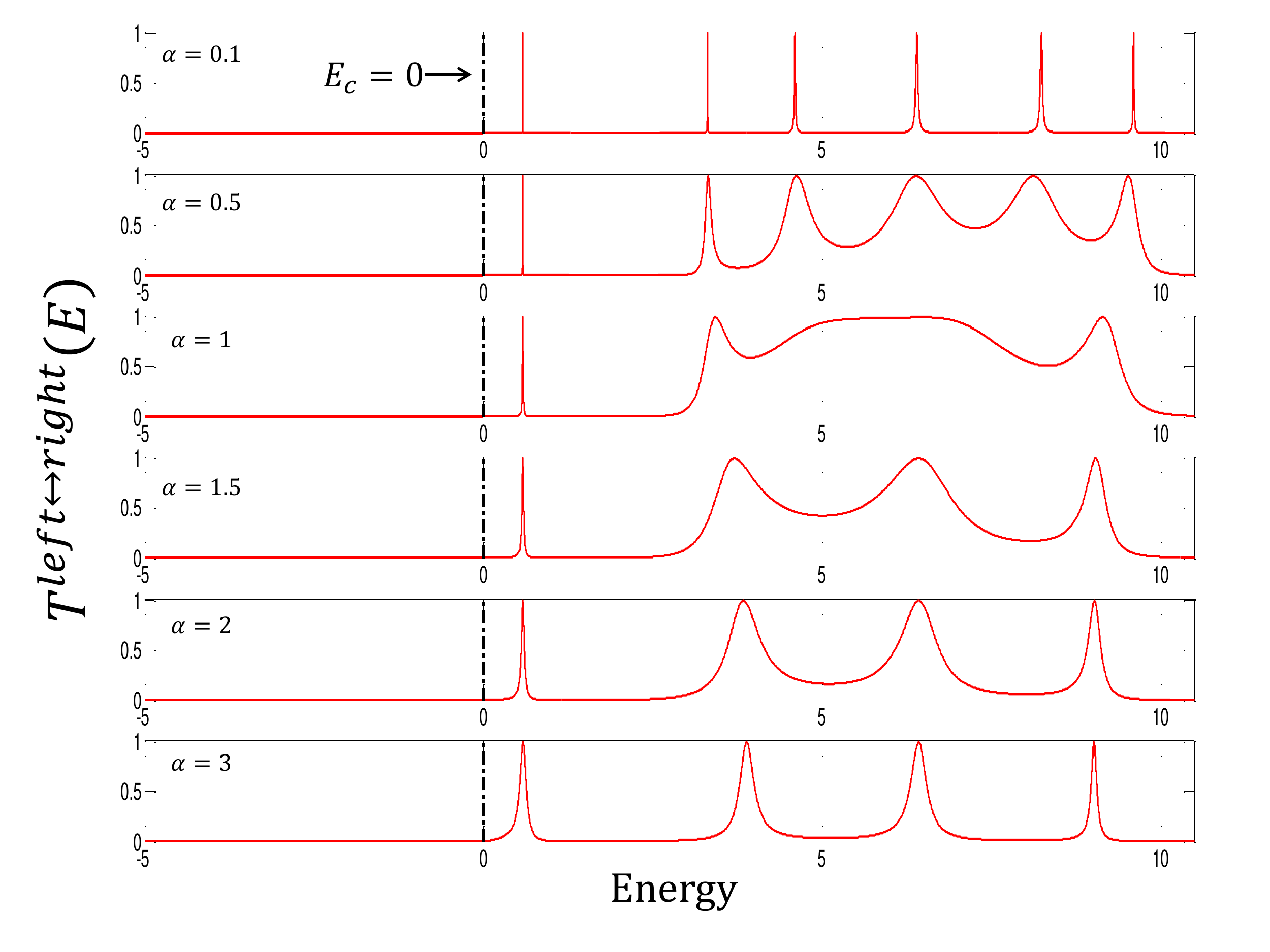}
\caption{(Color online). \small{Transmission through the two-channel wire with energy-dependent couplings for different values of the coupling strength, $\alpha$. The cutoff threshold located between the two bands is shown with a red vertical line, $E_{c}$=0. The wire parameters are the same as in Fig . \ref{fig:003}. }} \label{fig:007}
\end{figure}

In Fig. \ref{fig:007} we show the transmission coefficients for different values of the strength coupling $\alpha$. It is clear that the states below threshold do not participate in the transport of the signal. Similar to the energy-independent case, Fig. \ref{fig:005}, the transmission increases in the super-radiant transition when the resonances start overlapping.

\section{Conclusion}

The effective non-Hermitian Hamiltonian was used to study an open quantum wire of two-level atoms. The dynamics of the states when the wire is coupled to the continuum undergo dramatic change as the strength of the coupling to the continuum varies crossing the super-radiant segregation point separating the broad (short-lived) states from long-lived (trapped) states. It was demonstrated that the width distribution of the states significantly alters transport properties of the wire. Maximum transmission is achieved at the super-radiance transition. The future studies should include the practically important current-voltage characteristic of a quantum wire in different coupling regimes. This can be done considering
the wire connected with reservoirs of electrons with different changeable chemical potentials.
%%%%%%%%%%%%%%%%%%%%%%%%%%%%%%%%%%%%%%%%%%%%
%% Sample figure:
%%
%% The option [height=...] scales the picture to the given height,
%% without it it would be printed at its nominal size
%%%%%%%%%%%%%%%%%%%%%%%%%%%%%%%%%%%%%%%%%%%%

%%%%%%%%%%%%%%%%%%%%%%%%%%%%%%%%%%%%%%%%%%%%%%%%
%% BACKMATTER
%%%%%%%%%%%%%%%%%%%%%%%%%%%%%%%%%%%%%%%%%%%%%%%%

\begin{theacknowledgments}
A.T. is thankful to the organizers for the given opportunity to present the results at the conference.
V.Z. acknowledges the support through the NSF grants PHY-1068217 and PHY-1404442.
\end{theacknowledgments}



\bibliographystyle{aipproc}   % if natbib is available

%\bibliography{sample}

 % \typeout{******************************************}
 % \typeout{** Please run "bibtex \jobname" to optain}
 % \typeout{** the bibliography and then re-run LaTeX}
  %\typeout{** twice to fix the references!}
  %\typeout{******************************************}
  %\typeout{}
 %}

%%%%%%%%%%%%%%%%%%%%%%%%%%%%%%%%%%%%%%%%%%%
%% The following lines show an example how to produce a bibliography
%% without the help of the BibTeX program. This could be used instead
%% of the above.
%%%%%%%%%%%%%%%%%%%%%%%%%%%%%%%%%%%%%%%%%%%

%%
%% End of file `template-8s.tex'.
\end{document}